\documentclass{article}
\usepackage{spconf,amsmath,graphicx}
\usepackage{multirow,booktabs}
\usepackage{multicol}
\usepackage[justification=centering]{caption}
\usepackage{mathabx}
\usepackage[top=0.75in, bottom=1.1in, left=0.75in, right=0.75in]{geometry}
\usepackage{enumitem}
\usepackage{url}
\usepackage{hyperref} 
\usepackage{pdfpages}

\title{A comparative study of quality and content-based spatial pooling strategies in image quality assessment}

\name{Dogancan Temel and Ghassan AlRegib}

\address{Center for Signal and Information Processing (CSIP)\\
School of Electrical and Computer Engineering\\
Georgia Institute of Technology, Atlanta, GA, 30332-0250 USA\\
\{cantemel,alregib\}@gatech.edu}

\begin{document}

\onecolumn 

\begin{description}[labelindent=1cm,leftmargin=3cm,style=multiline]

\item[\textbf{Citation}]{D. Temel and G. AlRegib, "A comparative study of quality and content-based spatial pooling strategies in image quality assessment," 2015 IEEE Global Conference on Signal and Information Processing (GlobalSIP), Orlando, FL, 2015, pp. 732-736.
} \\

\item[\textbf{DOI}]{\url{https://doi.org/10.1109/GlobalSIP.2015.7418293}} \\

\item[\textbf{Review}]{Date added to IEEE Xplore: 25 February 2016} \\

\item[\textbf{Code/Slides}]{\url{https://ghassanalregib.com/publications/}} \\

\item[\textbf{Bib}] {
@INPROCEEDINGS\{Temel2015\_GlobalSIP, \\
author=\{D. Temel and G. AlRegib\},\\ 
booktitle=\{2015 IEEE Global Conference on Signal and Information Processing (GlobalSIP)\},\\ 
title=\{A comparative study of quality and content-based spatial pooling strategies in image quality assessment\},\\ 
year=\{2015\},\\ 
pages=\{732-736\},\\
doi=\{10.1109/GlobalSIP.2015.7418293\},\\ 
month=\{Dec\},\}\\
} \\

\item[\textbf{Copyright}]{\textcopyright 2015 IEEE. Personal use of this material is permitted. Permission from IEEE must be obtained for all other uses, in any current or future media, including reprinting/republishing this material for advertising or promotional purposes,
creating new collective works, for resale or redistribution to servers or lists, or reuse of any copyrighted component
of this work in other works. } \\

\item[\textbf{Contact}]{\href{mailto:alregib@gatech.edu}{alregib@gatech.edu}~~~~~~~\url{https://ghassanalregib.com/}\\ \href{mailto:dcantemel@gmail.com}{dcantemel@gmail.com}~~~~~~~\url{http://cantemel.com/}}
\end{description} 

\thispagestyle{empty}
\newpage
\clearpage

\twocolumn

%
\maketitle
\begin{abstract}
The process of quantifying image quality consists of engineering the quality features and pooling these features to obtain a value or a map. There has been a significant research interest in designing the quality features but pooling is usually overlooked compared to feature design. In this work, we compare the state of the art quality and content-based spatial pooling strategies and show that \texttt{although features are the key in any image quality assessment}, pooling also matters.  We also propose a quality-based spatial pooling strategy that is based on linearly weighted percentile pooling (\textbf{WPP}). Pooling strategies are analyzed for squared error, SSIM and PerSIM in LIVE, multiply distorted LIVE and TID2013 image databases.

\end{abstract}
\begin{keywords}
image quality assessment, quality/distortion maps, spatial pooling, statistical significance
\vspace{-0.2cm}
\end{keywords}
\vspace{-0.4cm}
\section{Introduction}
\label{sec:intro}
\vspace{-0.4cm}

Image quality models are designed to estimate the perceived quality of images. The design of the models requires engineered features that are correlated with the perceived quality. Moreover, extracted features need to be combined to obtain the quality estimate. There has been a significant effort in engineering image quality attributes most of which focused on the feature design part. However, pooling strategy selection is commonly overlooked and mean pooling is used in most of the quality estimators without further investigating alternative approaches.    

The authors in \cite{Wang2006} investigate the effect of spatial pooling strategies for pixel-wise and  structural image quality metrics. Minkowski pooling and local quality/distortion-weighted pooling are compared with the information content weighted pooling. In \cite{Moorthy2009}, the authors propose a pooling scheme based on the fact that significant degradation over the images dominate the perceived quality. A percentile pooling approach is followed where highly distorted regions are scaled before the fusion of the similarity map. The authors in \cite{Zewdie14} combine percentile information with mean, median and max values of the quality map to obtain the quality estimate. In \cite{Gong12}, the authors evaluate existing pooling strategies for color printing quality attributes, which can be sorted as sharpness, color, lightness, contrast and artifacts. In addition to the quality-based pooling, content-weighted pooling methods are also used in the comparison. There are also human visual system and fixation-based models in the literature but the scope of this paper is limited to the quality and content-based pooling strategies.

In this paper, we perform a comparative study of spatial pooling strategies in terms of linearity and ranking with statistical significance tests. We also propose a spatial pooling strategy based on the observation that the perceived quality is dominated by highly degraded regions. Instead of following a standard percentile pooling strategy, we calculate the quality thresholds that correspond to the percentile limits for various percentages. In case of  quality maps, these percentile values  are linearly weighted so that percentile limits close to $0.0\%$ get the highest weights whereas the ones close to $100.0\%$ are scaled with the lowest weights. The opposite scenario is valid in case of the distortion maps. Scaled percentile limits are summed and divided by the sum of the weights to obtain a normalized quality indicator.

\vspace{-0.45cm}

\section{Spatial Pooling Strategies}
\label{sec:main}
\vspace{-0.3cm}

In this section, we briefly describe quality- and content-based spatial pooling strategies used in the literature. Then, we introduce the weighted percentile pooling (\texttt{WPP}). 

\vspace{-0.4cm}

\subsection{Basic Statistics}
\label{main:basic}
\vspace{-0.2cm}

The most common way to map a distortion/quality map to a final value is by calculating the mean. Moreover, other basic statistical information including but not limited to standard deviation, median, min and max are also used in the pooling.

\vspace{-0.5cm}

\subsection{Percentile Pooling}
\label{main:percentile}
\vspace{-0.2cm}
Severe distortions dominate the perceived quality of images. Therefore, percentile-based methods  try to estimate the threshold that bounds the pixels with significantly perceivable degradation. These methods simply scale the distortion/quality values that fall into the target percentile. If we have a quality map, we focus on the low quality values and the quality values in the target percentile are divided with a scalar $c_1$ as expressed in Eq. (\ref{eq:perc_1}).   

\vspace{-0.5cm}

\begin{equation}
\label{eq:perc_1}
\hat{Q}[m,n] =\left\{ {\begin{array}{*{20}{l}}
Q[m,n] / c_1 ,&Q[m,n]< perc(Q,p)\\
Q[m,n], & \text{otherwise}
\end{array}} \right.
,
\vspace{-0.3cm}
\end{equation}

\hspace{-1.8em}
where $Q[m,n]$ is the quality map entry with the pixel locations $m$ and $n$, respectively. $perc(\cdot)$ is the percentile function that returns the percentile of the values in the map ($Q$) where $p$ is the parameter that corresponds to the the target percentage in the interval $[0,100]$. In case of the distortion maps, highly distorted entries are multiplied with the same constant. The authors in \cite{Moorthy2009} tune the percentile pooling metric for the structural similarity index where $p$ is set to $6.0$ and $c_1$ is set to $4000$ as a consequence of sweeping the parameter space with a step size of $1\%$ to find the configuration that corresponds to the highest Spearman correlation coefficient.  

\vspace{-0.4cm}

\subsection{5-Number Summary}
\label{main:5Point}
\vspace{-0.1cm}

The authors in \cite{Zewdie14} combine basic statistical information with percentile thresholds to obtain 5-Number summary in Eq. (\ref{eq:5Point}).

\vspace{-0.8cm}

\begin{equation}
\label{eq:5Point}
5_{Num}=\frac{mean(Q)+Q1+median(Q)+Q3+max(Q)}{5} 
,
\end{equation}
\vspace{-0.5cm}

\hspace{-1.8em}
where mean, median and max are calculated over the full resolution quality/distortion maps and Q1 and Q3 are equivalent to $perc(Q,25)$ and $perc(Q,75)$, respectively.	

\vspace{-0.5cm}

\subsection{Minkowski}
\label{main:minkowski}
\vspace{-0.2cm}

Minkowski-based pooling includes pixel-wise mapping of the quality/distortion maps using a power function as explained in Eq. (\ref{eq:Minkowski}).
\vspace{-0.7cm}

\begin{equation}
\label{eq:Minkowski}
Minkowski= \sum_{m=1}^{M} \sum_{n=1}^{N} \\
  \frac {Q[m,n]^p} {M \cdot N}
  ,\vspace{-0.2cm}
\end{equation}

\hspace{-1.8em}
where pixel indexes are denoted as $m$ and $n$ and the number of rows and columns are represented with $M$ and $N$. The most common $p$ values used in the literature are $1/8$, $1/4$, $1/2$, $2$, $4$ and $8$ so these values are used in the simulations. 
\vspace{-0.4cm}

\subsection{Quality/Distortion Weighted Pooling}
\label{main:QualityWeighted}

\vspace{-0.1cm}

Pixel values in the quality/distortion maps are weighted using a monotonic function and the weighted values are summed up over the full resolution map. Then, the obtained sum is divided by the sum of the weights for normalization as descibed in Eq. (\ref{eq:QualityWeighted}).

\vspace{-0.6cm}

\begin{equation}
\label{eq:QualityWeighted}
QW=  \frac {\sum_{m=1}^{M} \sum_{n=1}^{N} w[m,n] \cdot Q[m,n]} {\sum_{m=1}^{M} \sum_{n=1}^{N} w[m,n]}
  .\vspace{-0.1cm}
\end{equation}

\hspace{-1.8em}
The weight term is the $p^{th}$ power of the pixel-wise quality value as expressed in Eq. (\ref{eq:QualityWeighted_2}).

\vspace{-0.4cm}

\begin{equation}
\label{eq:QualityWeighted_2}
w[m,n]=Q[m,n]^p
\end{equation}

\vspace{-0.2cm}

\hspace{-1.8em}
 Distortion weighted pooling is obtained when quality map is replaced with a distortion map.

\vspace{-0.4cm}

\subsection{Information Weighted Pooling}
\label{main:IW}
\vspace{-0.2cm}

The methods defined in Sections \ref{main:basic}-\ref{main:QualityWeighted} are based solely on the quality/distortion maps. In addition to the information within these maps, the authors in \cite{Wang2006} propose using the additional information included in the reference and distorted images. Information content is quantified as the number of bits that can be received from an image that passes through a noisy channel. The source is assumed to follow a local Gaussian model and the channel characteristic is modeled with additive Gaussian to make the problem tractable. Information-based weighting is expressed in Eq. (\ref{eq:IW}).  

\vspace{-0.6cm}

\begin{equation}
\label{eq:IW}
w[m,n]=log \left[  \left( 1+ \frac {\sigma_I[m,n]^2} {c_2} \right) \left( 1+ \frac {\sigma_J[m,n]^2} {c_2} \right) \right]
,\vspace{-0.1cm}
\end{equation}

\hspace{-1.8em}
where $\sigma_{I}[m,n]$ is the standard deviation of the reference image and  $\sigma_{J}[m,n]$ is the standard deviation of the distorted image, and $c_2$ is a constant introduced to represent the channel noise. In our experiments, we simulate six different configurations of the information content weighted model. First, the distortion map is weighted with the expression given in Eq. (\ref{eq:IW}). We simulate the scenarios where either the reference or the distorted image information is used. A sliding window is used with two configurations, with and without Gaussian masking, which has a standard deviation of $1.5$ pixels.

\vspace{-0.4cm}

\subsection{WPP: Weighted Percentile Pooling}
\label{main:WPP}
\vspace{-0.1cm}
Standard percentile pooling is used to scale up the significance of highly distorted regions as described in Section \ref{main:percentile}. However, all pixels are scaled with a constant value instead of an adaptive mapping. Percentile thresholds can also be used to estimate quality as in Section \ref{main:5Point} but calculating the linear combination of basic statistics along with percentile thresholds are not very intuitive and lack of adaptation. In this work, pooling strategy is also based on percentile pooling where we  calculate percentile thresholds in between $1.0$ and $100.0$. Then, the percentile thresholds are scaled according to their relative significance with respect to percentile values because human visual system is more sensitive to severe degradations. Finally, percentile values are used for normalization. Weighted percentile pooling (\texttt{WPP}) can be used for quality maps as expressed in Eq. (\ref{eq:WPP_1}) as well as for distortion maps as in Eq. (\ref{eq:WPP_2}). The difference between quality and distortion pooling is based on the fact that low values in the quality map (Q) lead to significant degradation whereas they are the high ones in the distortion map (D). 
     
\vspace{-0.3cm}

\begin{equation}
\label{eq:WPP_1}
QW=  \frac {\sum_{s=1}^{T}  \left( 1- \frac{w_q(s)} {100} \right) \cdot perc\left(Q, w_q[s]  \right)} {\sum_{s=1}^{T}  \left( 1- \frac{w_q(s)} {100} \right)}.
\end{equation}

\vspace{-0.3cm}

\begin{equation}
\label{eq:WPP_2}
DW=  \frac {\sum_{s=1}^{T}  \left( \frac{w_d(s)} {100} \right) \cdot perc\left( D,w_d[s]  \right)} {\sum_{s=1}^{T}  \left(\frac{w_d(s)} {100} \right)}.
\end{equation}

\vspace{-0.2cm}

\hspace{-1.8em}
The weights of the quality and distortion maps are denoted as  $w_q$ and $w_d$, respectively, the term $s$ is an index that is based on the number of percentiles used in the combination, and $T$ is the upper limit of the index. 

\vspace{-0.7cm}

\begin{equation}
\label{eq:WPP_weight_Q}
w_q[s] =\left\{ {\begin{array}{*{20}{l}}
1+\frac{100}{N_{bin} \cdot} \cdot s, &1+\frac{100}{N_{bin} \cdot} \cdot s < 100\\
1, & \text{otherwise}
\end{array}} \right.
\end{equation}

\vspace{-0.5cm}

\begin{equation}
\label{eq:WPP_weight_D}
w_d[s] =\left\{ {\begin{array}{*{20}{l}}
100-\frac{100}{N_{bin} \cdot} \cdot s, &100-\frac{100}{N_{bin} \cdot} \cdot s > 1\\
100, & \text{otherwise}
\end{array}} \right.
\end{equation}

The number and value of the percentiles used in the pooling are based on the input ($N_{bin}$). In this work, we only use three configurations where $N_{bin}$ is set to $1$, $10$ or $20$. The term $s$ is defined over the range where the percentile values are greater than or equal to $1$ or less than or equal to $100$.    

\vspace{-0.4cm}

\section{Validation}
\label{sec:validation}
\vspace{-0.3cm}
Image databases with subjective scores are used to validate metric performance. In order to compare the pooling strategies, quality attributes are extracted using different assessment methods and then different pooling strategies are used to pool the attribute map into a final score to be compared with subjective scores. Squared error  between reference and distorted images, the SSIM \cite{Bovik2004} and the PerSIM \cite{Temel2015} are used as the quality attributes. Pearson correlation coefficient is used to measure linearity and Spearman correlation coefficient is used to perform ranking-based comparison. Monotonic logistic regression described in \cite{LIVE2006} is used  for a fair comparison in terms of linearity. 

\begin{figure}[htbp!]
        \vspace{-0.4cm}

\begin{center}
\noindent
  \includegraphics[width=0.40\linewidth]{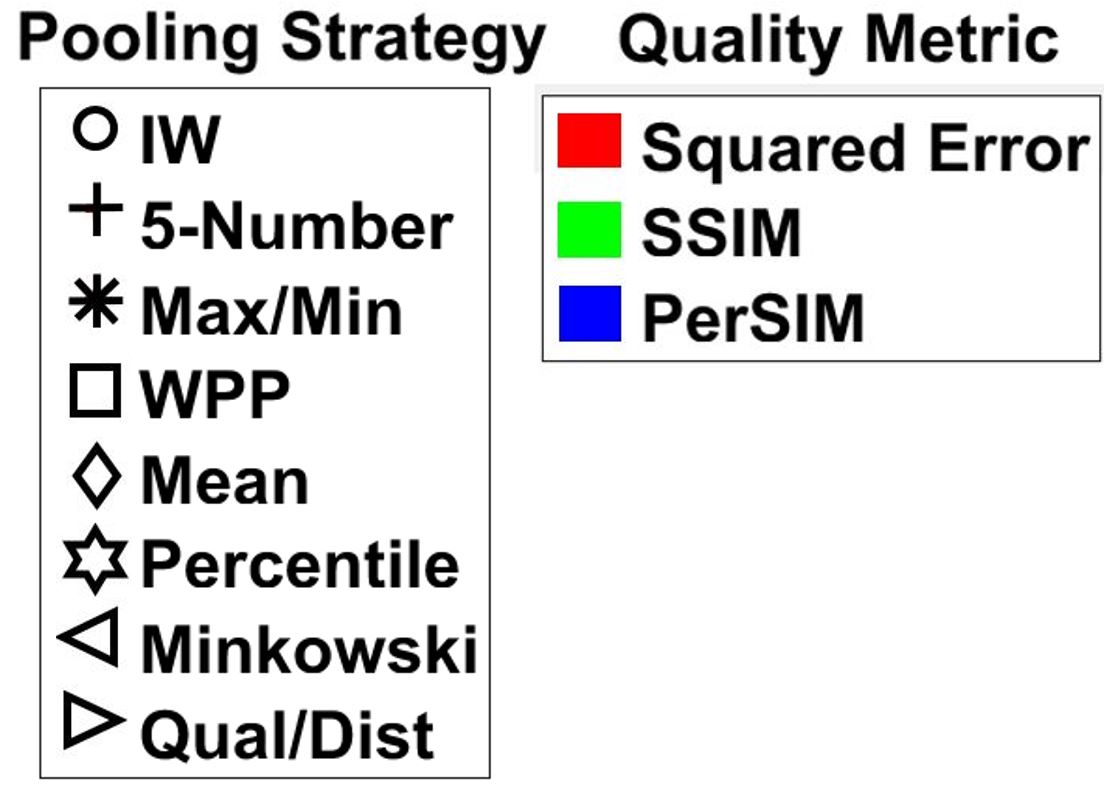}  
        \vspace{-0.28cm}

  \caption{Legend symbols and colors}
  \label{fig:legend}
      \vspace{-0.9cm}

\end{center}
\end{figure}

The pooling strategies described in Section \ref{sec:main} are used in comparison. Mean pooling is used for all the strategies, minimum is used for SSIM and PerSIM and maximum is used for squared error.
There is only one configuration reported for percentile and 5-Number pooling since their parameters are already tuned. In case of other pooling strategies, the best performing configuration is reported in each distortion type. The legend of the figures is given in Fig. \ref{fig:legend} where different pooling strategies are represented with different shapes and the type of the quality attributes is shown with different colors. The abbreviations are summarized as follow: MK: Minkowski, QD:Quality/Distortion weighted, IW: Information weighted, M/M: Min for quality and Max for distortion, 5-N.:5-Number summary, Per.:Percentile and WPP:Weighted percentile pooling. In the following figures, the vertical axis correponds to the correlation value and the horizontal axis corresponds to the distortion types.

\begin{figure}[htbp!]
      \vspace{-0.35cm}

\begin{center}
\noindent
  \includegraphics[width=0.6\linewidth, trim= 20mm 68mm 20mm 65mm]{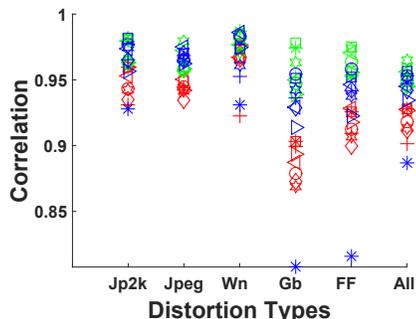}

        \vspace{-0.30cm}

  \caption{LIVE Database - Pearson CC}
    \label{fig:live_pearson}
      \vspace{-0.2cm}

\end{center}
\end{figure}

\begin{figure}[htbp!]
      \vspace{-0.9cm}

\begin{center}
\noindent
  \includegraphics[width=0.6\linewidth, trim= 20mm 68mm 20mm 65mm]{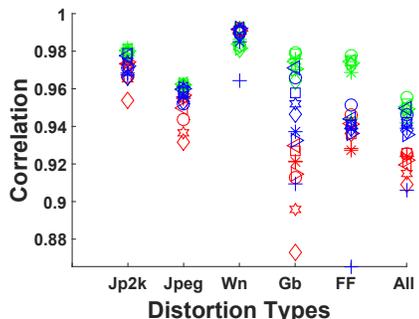}        \vspace{-0.30cm}

  \caption{LIVE Database - Spearman CC}
  \label{fig:live_spearman}
      \vspace{-0.8cm}

\end{center}

\end{figure}

The performance of the pooling strategies in the LIVE database \cite{LIVE2006} are summarized in Fig.\ref{fig:live_pearson} (Pearson) and Fig.\ref{fig:live_spearman} (Spearman). The performance of the metrics in case of compression artifacts (Jp2k, Jpeg) are close to each other.  Max pooling and 5-point pooling using PerSIM and 5-number pooling using MSE perform worse compared to other pooling strategies in case of white noise (Wn). In case of Gaussian blur (Gb) , weighted percentile, max and information weighted pooling are the best performing pooling strategies. Weighted percentile, percentile and information weighted pooling outperform other strategies under fastfading distortions (FF). Overall, percentile  and weighted percentile pooling lead to highest Pearson correlation and in terms of Spearman correlation, information weighted and weighted percentile pooling using SSIM are the best ranked strategies.

\begin{figure}[htbp!]
\vspace{-0.35cm}

\begin{center}
\noindent
  \includegraphics[width=0.6\linewidth, trim= 20mm 68mm 20mm 65mm]{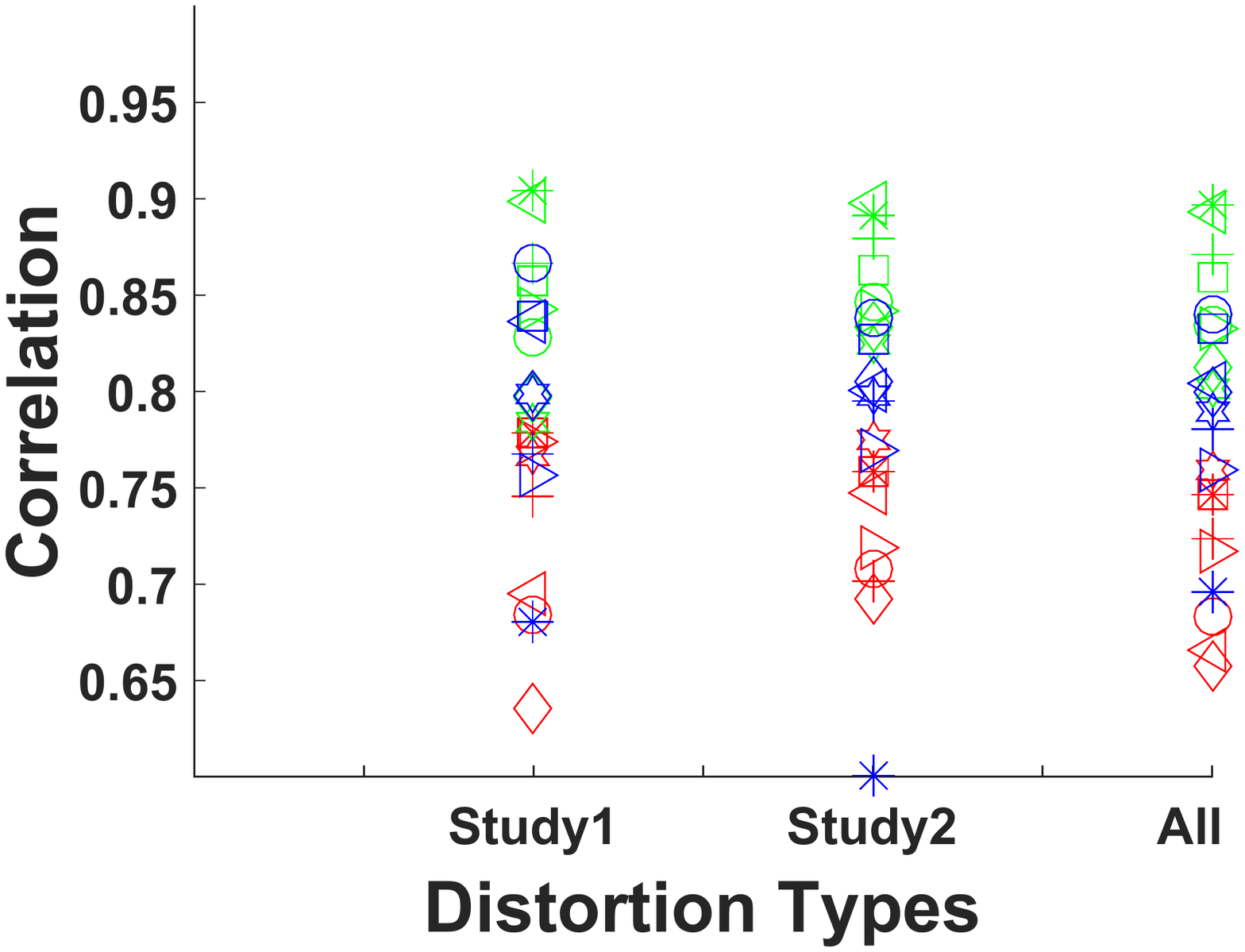}

        \vspace{-0.30cm}

  \caption{Multi Database - Pearson CC}
    \label{fig:multi_pearson}
      \vspace{-0.80cm}

\end{center}
\end{figure}

\begin{figure}[htbp!]
\vspace{-0.3cm}
\begin{center}
\noindent
  \includegraphics[width=0.6\linewidth, trim= 20mm 68mm 20mm 65mm]{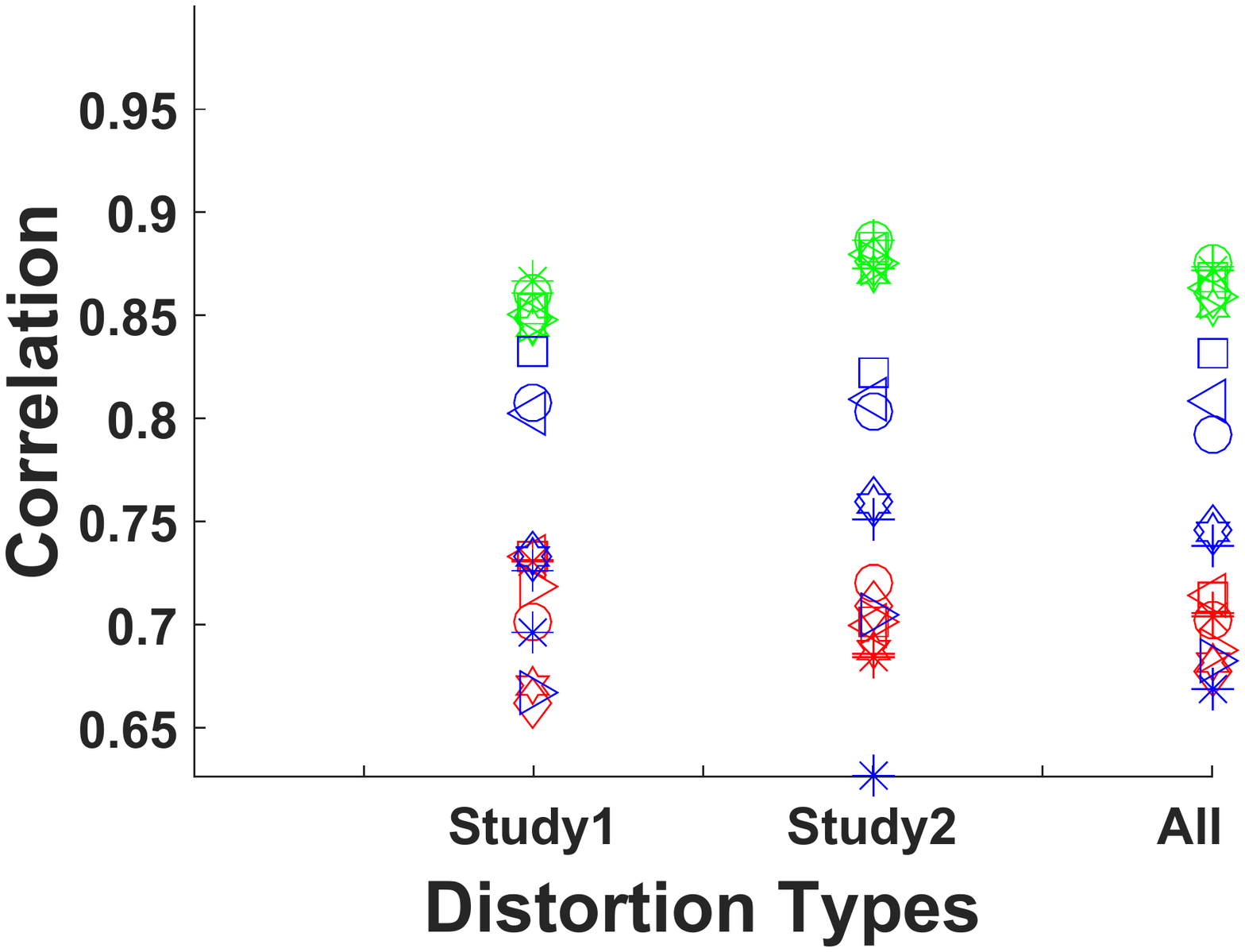}        \vspace{-0.25cm}

  \caption{Multi Database - Spearman CC}
  \label{fig:multi_spearman}
  \vspace{-0.45cm}

\end{center}

\end{figure}

In the multiply distorted LIVE database \cite{Jayaraman12}, Minkowski and Max/Min corresponds to highest linearity for pooling SSIM maps whereas information-weighted and weighted percentile pooling are the highest while using PerSIM as shown in Fig.\ref{fig:multi_pearson}. Percentile pooling is the most linear estimator while using squared error maps. In terms of ranking, information weighted and max/min are the highest for SSIM and weighted percentile pooling is the best for PerSIM whereas different strategies lead others in different categories in squared error metric as given in Fig.\ref{fig:multi_spearman}. 

\begin{figure}[htbp!]
\begin{center}
\noindent
  \includegraphics[width=0.6\linewidth, trim= 20mm 68mm 20mm 65mm]{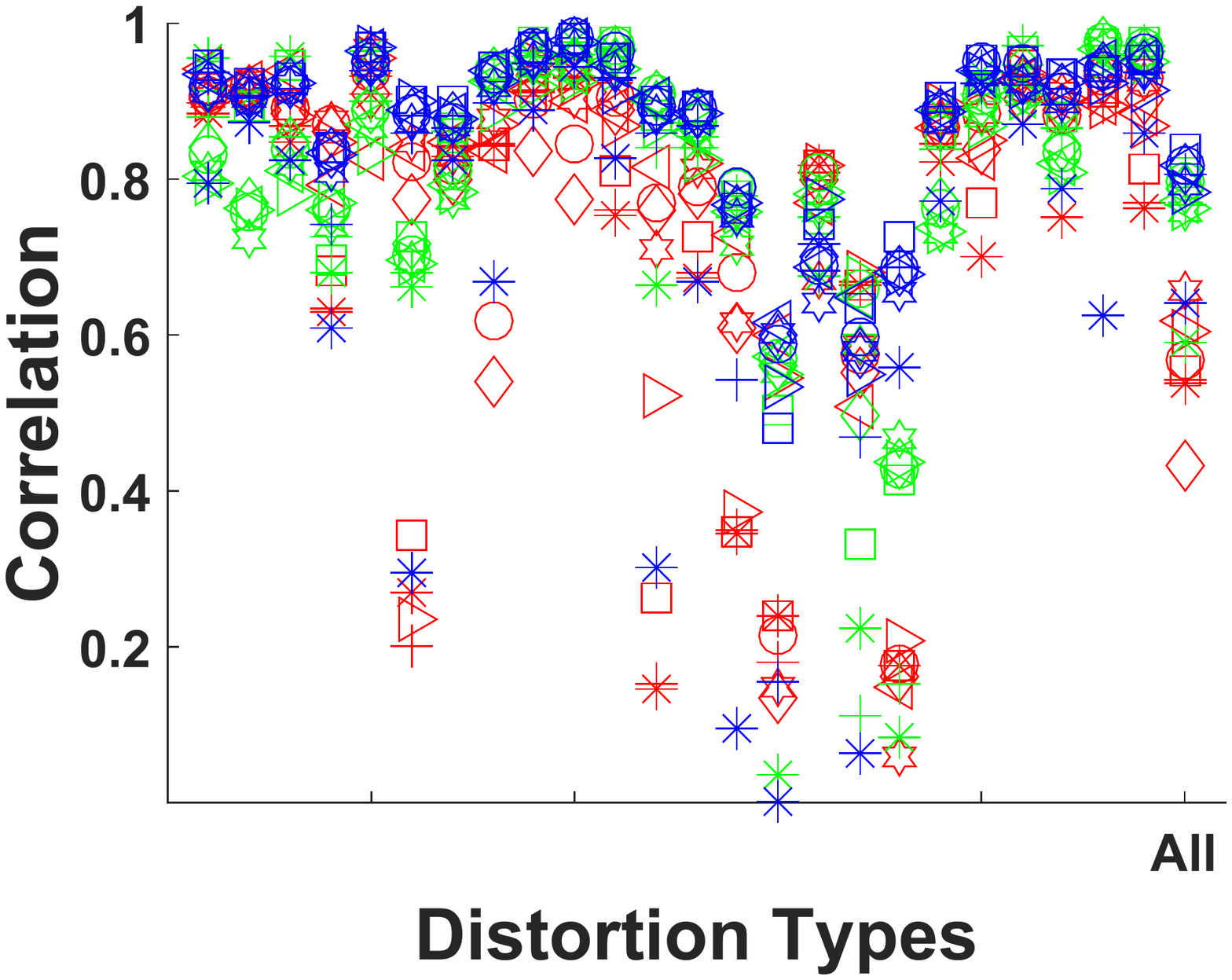}

        \vspace{-0.28cm}

  \caption{TID2013 Database - Pearson CC}
    \label{fig:tid_pearson}
      \vspace{-0.9cm}

\end{center}
\end{figure}

\begin{figure}[htbp!]
\vspace{-0.20cm}
\begin{center}
\noindent
  \includegraphics[width=0.6\linewidth, trim= 20mm 68mm 20mm 65mm]{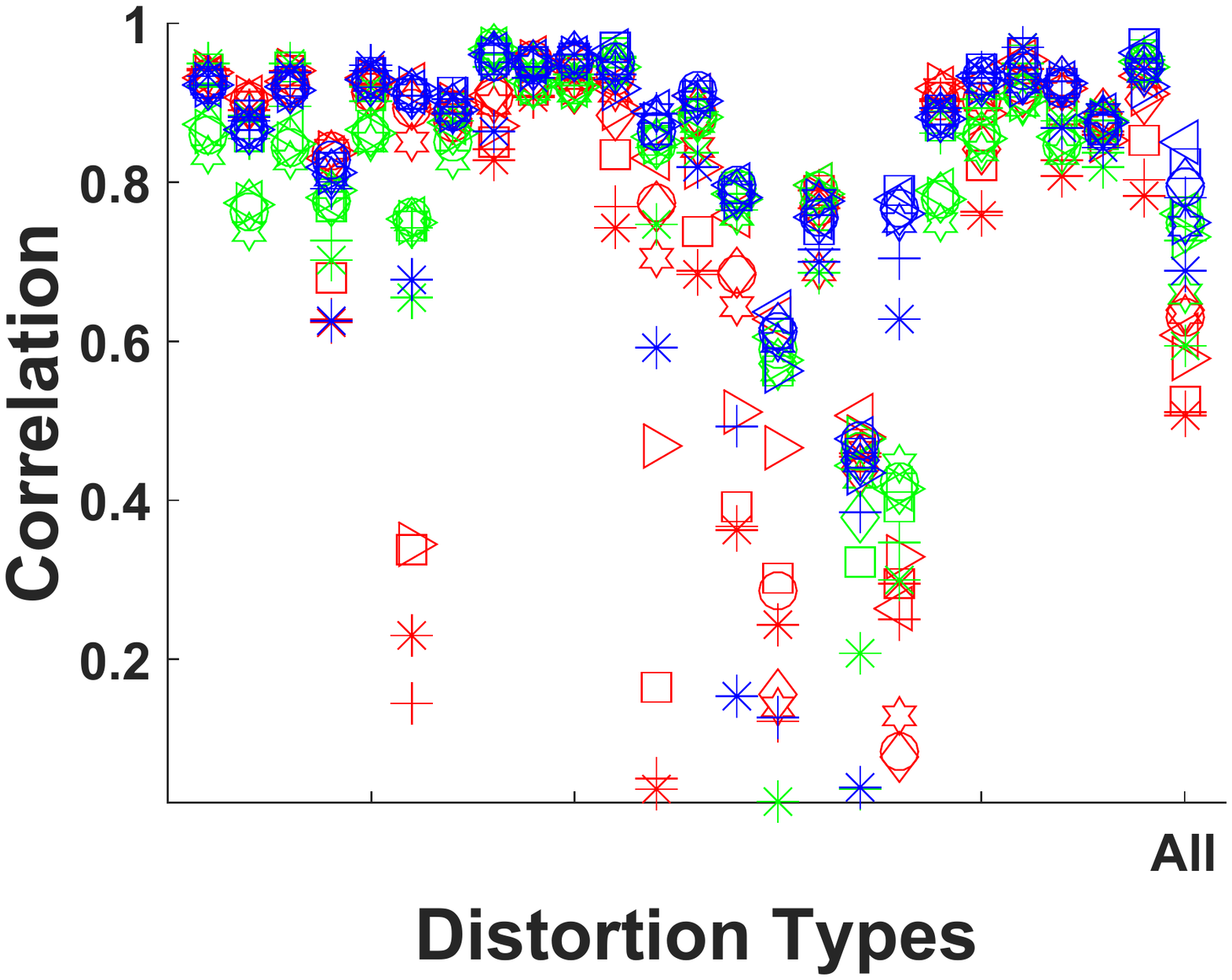}        \vspace{-0.28cm}

  \caption{TID2013 Database - Spearman CC}
  \label{fig:tid_spearman}
      \vspace{-0.7cm}

\end{center}

\end{figure}

TID image database \cite{TID2013} consists of $24$ different distortion types and the performance of the pooling strategies in each distortion category is shown in Fig.\ref{fig:tid_pearson} and Fig.\ref{fig:tid_spearman}. Because of the space constraints, we only discuss the performance of the strategies in the overall databases. In terms of linearity, weighted percentile pooling using PerSIM leads to the best performance followed by various strategies that are very close to each other. Minkowski-based pooling using PerSIM is the best in terms of ranking followed by weighted percentile pooling using PerSIM.

\begin{figure*}[htbp!]
\begin{center}
\noindent
  \includegraphics[width=\linewidth]{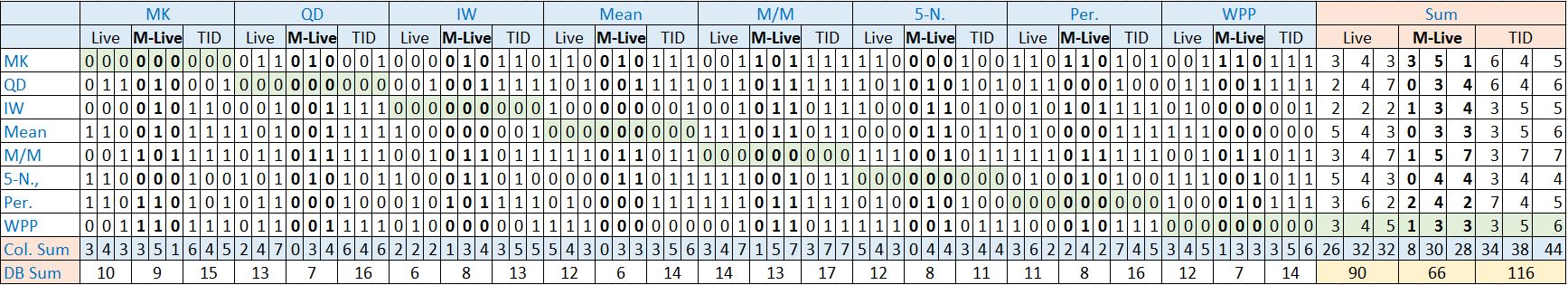}
            \vspace{-0.7cm}

  \caption{Significance of the differences between the Pearson correlation coefficients for various pooling strategies using different quality attributes and databases}
    \label{tab:stats}
          \vspace{-1.2cm}

\end{center}
\end{figure*}

Correlation coefficients are used to compare the performance of the pooling strategies. However, we also need to analyze the statistical significance of the differences to claim that the decrease or increase in the correlation actually matters.  We use the significance of the difference between the correlation coefficients suggested in ITU-T Rec. P.1401 \cite{ITU1401}. Statistical significance test results are summarized in Fig. \ref{tab:stats}. The results of each pooling strategy is a nine digit codeword where the first three corresponds to the LIVE database, the second three is multiply disorted LIVE (M-Live) and the third tree corresponds to TID2013 database (TID). In these ternary groups, the first attribute is squred error, the second is SSIM and the third is PerSIM. A 1 in the codeword means that there is significant difference between the pooling strategies highlighted in the row and column titles in terms of Pearson correlation coefficient otherwise the difference is insignificant.

Statistical significance tests show that none of the pooling strategies is different than the others in all databases and metric types in terms of Pearson correlation. Database selection is important in comparing the pooling strategies. The sum of the individual columns (Col. Sum) corresponds to the total statistical difference in a specific database using a specific attribute.  For each quality attribute, TID leads to the highest statistical significance total since it has a wide range of distortion types where different pooling strategies can stand out. When the statistical significance totals are summed up for each database (DB Sum), we can see that multiply distorted LIVE database (M-Live) has the least statistically different comparisons (66) and TID has the most (116).

 The proposed pooling strategy weighted percentile pooling (\texttt{WPP}) is inherently designed to calculate the percentile over quality maps that contain perceivable degradation so \texttt{WPP} is more consistent in structural and  perceptual similarity-based  pooling.  However, in case of pixel-wise error pooling, the accuracy of \texttt{WPP} depends on the distortion type. In general, structural similarity  (SSIM) and perceptual similarity metrics (PerSIM) outperform pixel-wise difference with several exceptions.

      \vspace{-0.1cm}

\section{Conclusion}
      \vspace{-0.3cm}

\label{subsec:conc}
In the image quality assessment literature, mean pooling is commonly used to map the distortion/quality maps to the final value. However, when the performance comparison is considered, mean pooling is not consistent over different databases, distortion types and quality attributes. According to the comparison of spatial pooling strategies, we conclude that pooling strategies influence the performance of the estimator. However, feature selection is still more dominant in the accuracy of the final quality score. It is easier to differentiate the performance of pooling strategies and quality attributes as the number of distortion types increase in the validation set.

\newpage
	

\end{document}